\documentclass[prb,twocolumn,showpacs]{revtex4}

\usepackage{graphicx}
\usepackage{dcolumn}
\usepackage{bm}

\begin{document}


\title{Magnetization Reversal in Elongated Fe Nanoparticles}

\author{Yongqing Li}
\email{yqli@physics.ucsb.edu}
\author{Peng Xiong}
\author{Stephan von Moln\'{a}r}
\affiliation{MARTECH and Department
of Physics, Florida State University, Tallahassee, FL 32306-4351}
\author{Yuzo Ohno}
\author{Hideo Ohno}
\affiliation{Laboratory for Nanoelectronics and Spintronics,
Research Institute of Electrical Communication, Tohoku University,
Sendai, Japan}

\date{\today}

\begin{abstract}

Magnetization reversal of individual, isolated high-aspect-ratio
Fe nanoparticles with diameters comparable to the magnetic
exchange length is studied by high-sensitivity submicron Hall
magnetometry. For a Fe nanoparticle with diameter of 5\,nm, the
magnetization reversal is found to be an incoherent process with
localized nucleation assisted by thermal activation, even though
the particle has a single-domain static state. For a larger
elongated Fe nanoparticle with a diameter greater than 10\,nm, the
inhomogeneous magnetic structure of the particle plays important
role in the reversal process.

\end{abstract}

\pacs{75.75.+a, 85.75.Nn, 75.60.Jk, 75.60.Ch}

\maketitle

\section{Introduction}
Magnetic nanowires still attracted much attention despite a long
history of the study of elongated magnetic nanoparticles
\cite{Sellmyer01}. Besides some potential technological
applications such as high density magnetic information storage,
magnetic nanowires also provide a unique arena for testing
theoretical models of magnetization reversal. Recent work on
magnetization reversal \cite{Wernsdorfer96, Wegrowe99} and domain
wall motion \cite{Pietzsch01, Cayssol04} in magnetic nanowires has
provided in-depth understanding of magnetism on the nanometer
scale and even atomic scales.  In a magnetic nanowire, the
competition between the magnetostatic energy and magnetic exchange
interaction gives rise to a characteristic length, the coherence
diameter, $d_c\simeq7.3l_{ex}=7.3(A/4\pi M_s)^{1/2}$, where
$l_{ex}$ is the magnetic exchange length,  $A$ the exchange
constant and $M_s$ the saturation magnetization \cite{Exlength}.
For Fe, Co, and Ni, $d_c$ is about 11, 15 and 25\,nm, respectively
\cite{Sellmyer01}. For an ideal nanowire with diameter $d\ll d_c$,
micromagnetic theory predicts a coherent rotation of the entire
magnetic volume due to the dominance of exchange interactions,
while incoherent reversal modes such as curling are favored for
nanowires with $d>d_c$ \cite{Aharoni96}. A large body of
experimental results on magnetic nanowires have been analyzed in
terms of these fundamental reversal modes. Most of the
measurements, however, were carried out on ensembles of magnetic
nanowires, in which data analysis is complicated by the
distribution of size, shape, and microstructure of the nanowires,
as well as the dipolar interaction between the nanowires. In
addition, some important properties of individual particles such
as the stochastic nature of magnetization reversal are averaged
out. Measurements on single magnetic nanowires have only been
reported in a few cases. For example, electrodeposited Ni or Co
nanowires with diameters between 30\,nm and 90\,nm were studied
with micro-SQUID magnetometers \cite{Wernsdorfer96} and
subsequently by utilizing the anisotropic magnetoresistance (AMR)
effect \cite{Wegrowe99, Pignard00}. Experimental data from these
measurements have been fitted to the curling mode, but difficulty
arises due to the fact that the anisotropy energy from
magnetocrystalline and/or other effects are comparable to the
shape anisotropy \cite{Wegrowe99, Pignard00}. Hence, the study of
individual nanowires with one dominant anisotropy could provide
further insight into the fundamental magnetization reversal
process in such magnetic nanostructures. The Fe nanowires are one
of the ideal systems because their shape anisotropy is at least
one order of magnitude larger than the magnetocrystalline and
strain anisotropies \cite{Morrish65, Sokolov02}.

In this paper, we report the first detailed study of
\emph{individual} cylinder-shaped Fe nanoparticles with diameters
close to the coherence diameter. Measurements of a single Fe
nanocylinder with $d\simeq5$\,nm reveals the incoherent and
localized nature of the magnetization reversal although the
particle has a single-domain static state. We also present the
results on an elongated Fe nanoparticle with more structural
imperfections, as the first step toward studying more complex
magnetic dynamics in nanowires with diameters on the 10\,nm scale.

\section{Experiment}

Cylinder-shaped Fe nanoparticles were fabricated with a scanning
tunneling microscopy (STM) assisted chemical vapor deposition
technique, in which a precursor, Fe(CO)$_5$, was decomposed by
applying a high bias voltage to the STM tip \cite{McCord90,
Kent93}. Growth was controlled by the STM feedback electronics
operated in constant tunneling current mode. The fabricated Fe
nanoparticles have a polycrystalline bcc iron core surrounded by
an amorphous carbon coating \cite{Kent93}, which effectively
prevents the sample from oxidation \cite{Wirth99}. The diameter of
the iron core can be varied from a few nm to 20\,nm by varying the
growth rate, which can be controlled by the precursor pressure
\cite{Kent93, Wirth01}. The height of the Fe particles can be
precisely controlled by withdrawing the piezoelectric scanning
tube after it reaches a predetermined setpoint and can be varied
from tens of nanometers to a few microns \cite{Kent93}.  In order
to measure a single Fe nanoparticle, we grew a small number of Fe
nanoparticles with very large spacing onto a submicron Hall
magnetometer fabricated from a GaAs/AlGasAs heterostructure. The
heterostructure is consisted of a 1\,$\mu$m undoped GaAs layer,  a
30\,nm Al$_{0.33}$Ga$_{0.67}$As spacer, a 80\,nm n-doped
Al$_{0.33}$Ga$_{0.67}$As layer, and a 5\,nm GaAs cap. The carrier
concentration and the mobility of the two-dimensional electron gas
(2DEG) in this heterostructure are $n=2.2\times 10^{11}$\,cm$^2$
and $\mu=1\times10^5$\,cm$^2$\,/V$\cdot$s, respectively, which
were measured in the dark at $T=77$\,K. The STM-assisted growth
technique offers the critical high precision positioning
capability which makes the measurement of single nanoparticles
possible.

\begin{figure}
\includegraphics[width=6.5cm]{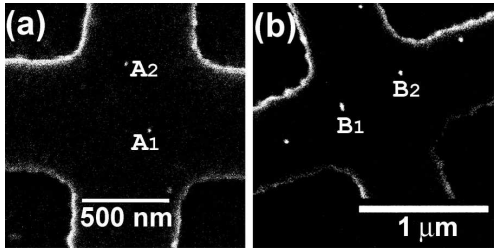}
\caption{\label{fig1} SEM images of two samples suitable for
measurements of single cylinder-shaped Fe nanoparticles with
$h\simeq120$\,nm. The stray field from the Fe particles is sensed
by 0.6$\times$0.6\,$\mu$m$^2$ GaAs/AlGaAs Hall magnetometers. (a)
The growth time for particles $A_1$ and $A_2$ is 11\,s. The
diameter of particle $A_1$ is estimated to be 5\,nm. (b) The
growth time for particles $B_1$ and $B_2$ are 42\,s and 35\,s,
respectively.}
\end{figure}

Shown in Fig.\,1 are scanning electron microscopy (SEM) images of
two samples to be presented in this paper. The height of the
particles was set to be $h\simeq120$\,nm, so the easy axis of the
particles is nearly perpendicular to the 2DEG plane due to the
dominance of shape anisotropy ($h\gg d$). The particles in sample
(a) were grown at a faster rate than those in sample (b), so the
diameters of particles $A_1$ and $A_2$ are expected to be smaller
than those of particles $B_1$ and $B_2$ \cite{Kent93}. The size of
the Hall crosses in both samples, which were fabricated with
electron beam lithography followed by wet chemical etching, is
about $0.6\times0.6$\,$\mu$m$^2$. By proper gating of the 2DEG,
and using a gradiometry setup in which the differential Hall
voltage between an empty Hall cross and a Hall cross with
nanoparticles grown on top is measured \cite{Kent94, Li04}, moment
sensitivity of submicron magnetometers has been improved to better
than $10^4$\,$\mu_B$/Hz$^{1/2}$ at 1\,Hz in a large applied
magnetic field \cite{Li03b}. The advantages of Hall magnetometry
over other single particle measurement techniques include the wide
range of operational temperature, no limitation in applied fields,
as well as the non-invasive nature of the measurement.

\section{Results and Discussion}

\begin{figure}
\includegraphics[width=7cm]{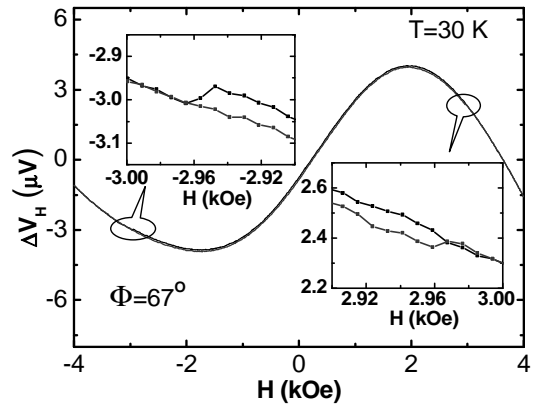}
\caption{\label{fig2} A typical hysteresis curve of sample (a)
from the Hall gradiometry measurement at $T=30$\,K with magnetic
filed applied 67\,$^\circ$ relative to the easy axis. The two
insets are the close-up view of magnetization reversals of the
particle $A_1$.}
\end{figure}

Hundreds of Hall measurements have been performed on sample (a) at
various temperatures and applied field angles (Earlier results on
this sample with field applied parallel to the easy axis of the Fe
nanoparticle were presented in Ref.\,\cite{Li02}).  In each of the
measurements, only one sharp jump in the Hall signal corresponding
to the switching of one particle is detected. Fig.\,2 shows a
typical hysteresis curve (raw data) measured at $T=30$\,K and
$\Phi=67$\,$^\circ$, where $\Phi$ is the angle between the applied
magnetic field and the easy axis of the Fe nanoparticle. The large
nonlinear background originates from the mesoscopic effects in the
small 2DEG structure at low temperatures. The observation of the
switching of one particle instead of two is not surprising: from
simple calculations the Hall signal from particle $A_1$ is
estimated to be about one order of magnitude larger than that of
particle $A_2$. In fact, the signal from particle $A_2$ is below
the noise level and thus undetectable in our measurements. When
the magnetic field is applied parallel to the easy axis, the net
hysteresis curve, obtained by subtracting the nonlinear
background, has a nearly rectangular shape \cite{Li02}. From the
Hall signal at zero field, the magnetic moment of the particle is
extracted from its average stray field over the active area of
Hall cross ($\sim0.4\times 0.4$\,$\mu$m$^2$). The diameter of
particle $A_1$ is estimated to be $5\pm1$\,nm, which is much
smaller than the coherence diameter of Fe nanowires (11\,nm).  The
dominance of exchange interaction at this length scale favors a
single domain structure. Indeed, we found that the hysteresis
curve of this particle is independent of field sweep history as
long as the field is swept from above the switching fields,
suggesting that this cylinder-shaped particle is a single-domain
particle at least by the static definition \cite{Monzon99}.

The magnetization reversal of a perfect nanowire with $d<d_c$ can
be described by coherent rotation. The corresponding angular
dependence of the switching fields is well described by the
Stoner-Wohlfarth model \cite{Stoner48}, and can be written as
$H_{sw}=H_K(\sin^{2/3}\Phi +\cos^{2/3}\Phi)^{-3/2}$, where $H_K$
is the anisotropy field. As shown in Fig.\,3, the measured angular
dependence of the switching fields of particle $A_1$ deviates
significantly from that expected from coherent rotation. The
switching fields at lower angles are much lower than those for
coherent rotation.  At high angles, the switching fields decrease
rapidly with decreasing $\Phi$, which is similar to previous
results on Fe nanoparticle arrays with $d=7$-14\,nm \cite{Li03a,
Wirth99}. Nonetheless, we notice that the switching field at
$\Phi\sim25$\,$^\circ$ appears to be slightly lower than that at
$\Phi\simeq0$\,$^\circ$, which might be regarded as a weak
signature of coherent rotation suggested in
Ref.\,\cite{Wernsdorfer96} for similar effects in Ni nanowires.
Such a feature was not observed in our previous measurements on Fe
nanoparticle arrays \cite{Wirth99, Li03a}.

\begin{figure}
\includegraphics[width=7.5cm]{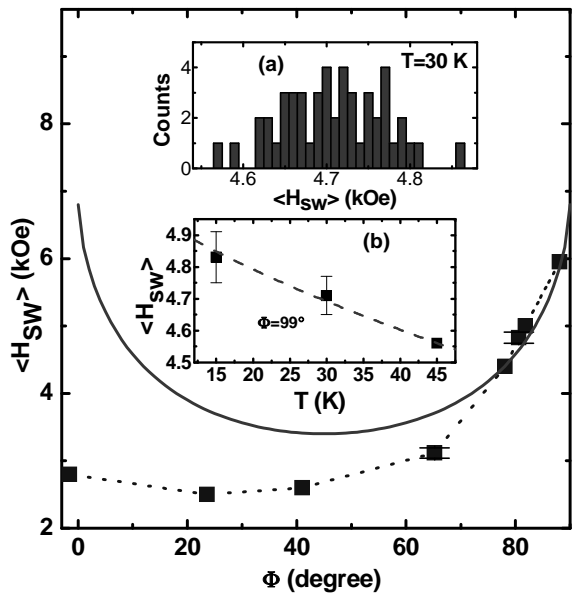}
\caption{\label{fig3} The angular dependence of the average
switching fields, $\langle H_{sw}\rangle$, of particle $A_1$ at
$T=30$\,K. Inset\,(a) shows the wide distribution of the switching
fields at $\Phi=99$\,$^\circ$ and $T=30$\,K. Inset\,(b) is the
temperature dependence of  $\langle H_{sw}\rangle$ at
$\Phi=99$\,$^\circ$. The spreading width of the switching fields
is shown by the height of the bars in inset\,(b), which is the
statistical standard deviation of $H_{sw}$ obtained from the
histogram measurements.}
\end{figure}
The deviation from coherent rotation is further manifested in the
histogram measurements. The stochastic nature of the switching
process causes the spreading of the switching fields in a certain
range. In the N\'{e}el-Brown model \cite{Neel49, Brown63}, the
magnetization reversal of a dynamic single domain particle at
finite temperature is described as thermal activation over a
single energy barrier. Kurkij\protect{\"{a}}rvi developed a
formalism mathematically connecting the switching statistics to
the energy barrier \cite{Kurkijarvi72}. Measurements on individual
ellipsoidal Co nanoparticles of $d=25$\,nm by Wernsdorfer et al.\
provided the first experimental agreement with the N\'{e}el-Brown
model \cite{Wernsdorfer97}. This model predicts an increase of the
spreading width and a decrease in the average switching field
$\langle H_{sw}\rangle$ with increasing temperature. This is not
the case for particle $A_1$. As shown in the inset of Fig.\,3, the
spreading width does not increase with temperature. The spreading
width estimated from Kurkij\protect{\"{a}}rvi's formulas is at
least one order of magnitude smaller than our experimental value
($\sim$2$\%$ of the $H_{sw}$ at T=15 K). Similar wide distribution
of $H_{sw}$ has also been observed in electrodeposited Ni
nanowires with d=75\,nm \cite{Wernsdorfer96}. The observed
$\langle H_{sw}\rangle$ does decrease with temperature as expected
from thermal activation, but it decreases much faster than the
prediction of the N\'{e}el-Brown model. The extracted energy
barrier, $E_0$, is $3\times10^4$\,K from the data at
$\Phi=99$\,$^\circ$ and $T=15$-45\,K (inset b in Fig.\,3),
corresponding to a thermal activation volume of
$\sim5\times10^2$\,nm$^3$, which is only about 1/5 of the total
volume of the nanocylinder. This is consistent with previous
magnetic viscosity measurements on large arrays of similar Fe
nanoparticles, in which the thermal activation volume was also
found to be a small fraction of the particles' total volume
\cite{Wirth01}. Similar results were also reported on
electrodeposited Ni nanowires \cite{Wernsdorfer96}.

The data presented above clearly suggest that the magnetization
reversal of particle $A_1$ is an incoherent process in which the
nucleation is localized and thermally assisted, despite the fact
that this particle has a diameter smaller than the coherence
diameter.  Recent micromagnetic simulation of defect-free d=9 nm
Fe nanopillars by Brown et al.\,\cite{Brown01, Brown04} showed
that nucleation can start at both ends and propagate through the
whole pillars in the time scale of nanoseconds. The calculated
$H_{sw}$ is about 2\,kOe for the fields applied parallel to the
easy axis, which is close to our experimental value. Another
theoretical work by Skomski et al.\ found that disorder such as
structural imperfections favor the localized magnetization
reversal over the delocalized modes such as coherent rotation and
curling \cite{Skomski00}.  The Fe nanoparticles fabricated with
STM are polycrystalline \cite{Kent93}, so multiple nucleation
sites including both the ends and the imperfections (e.g.\ grain
boundaries) are probably involved in the switching process. On the
other hand, the single-domain static state of this particle
suggests that these defects are still not strong enough to serve
as pinning centers for domain walls, which are overwhelmed by
magnetic exchange between strongly interacting magnetic clusters
(grains) favoring a uniform spin distribution. It will be of
particular interest to study how these structural imperfections
affect the propagation process after the nucleation, which may
determine the ultimate speed of spintronic devices based on such
elongated nanoparticles or similar nanostructures.

\begin{figure}
\includegraphics[width=7.5cm]{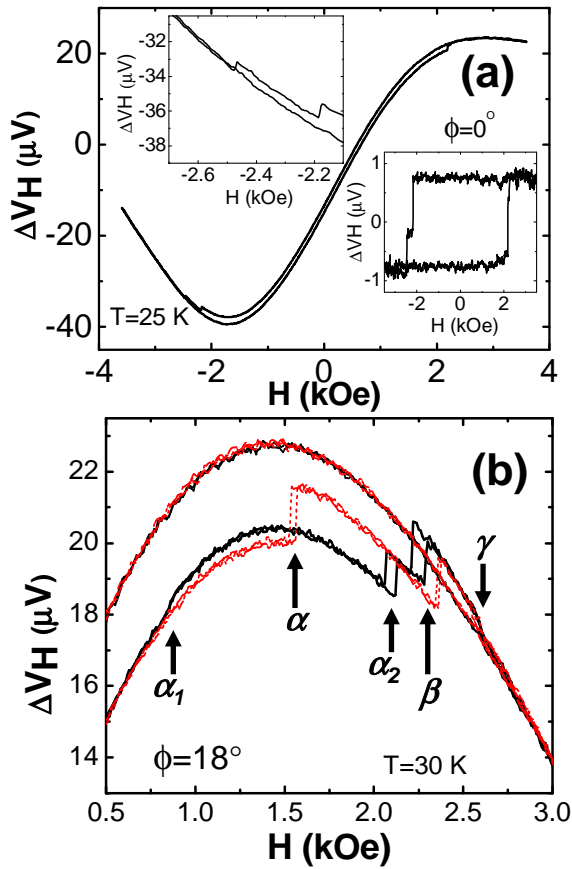}
\caption{\label{fig4}(Color online) (a) The hysteresis curve of
the sample (b) at $T=25$\,K and with magnetic field applied
perpendicular to the substrate. The upper-left inset is the
close-up view of the magnetic switching. The lower-right inset
shows the net hysteresis curve after the subtraction of non-linear
background. (b) Part of four hysteresis loops of the same sample
taken at $T=30$\,K and with fields applied 18\,$^\circ$  relative
to the normal of the substrate, i.e. $\phi=18$\,$^\circ$. Two
loops (black lines) have four switching events ($\alpha_1$,
$\alpha_2$, $\beta$, and $\gamma$), while the other two (red (dark
gray) lines) have three ($\alpha$, $\beta$, and $\gamma$). }
\end{figure}
Recently spintronic devices based on manipulation of domain walls
have been proposed \cite{Allwood02}. A reduction of domain wall
velocity with decreasing feature size has been reported recently
in submicron stripes of ultrathin Pt/Co/Pt films \cite{Cayssol04}.
As a first step toward the study of domain wall motion on the
10\,nm scale, we now demonstrate that elongated Fe nanoparticles
with domain walls can be fabricated with the STM-CVD technique.
Fig.\,4(a) shows the hysteresis curve of the sample (b) taken at
$T=25$\,K and with the field applied nearly parallel to the easy
axis, in which only two particles ($B_1$ and $B_2$) are close to
the active region of the Hall cross. After subtraction of the
nonlinear background, we obtain a rectangular shaped net
hysteresis curve, shown in the lower-right inset of Fig.\,4(a),
which exhibits two distinct sharp jumps in the Hall signal.  We
attribute these two jumps to the magnetization reversal of the
particle $B_1$ alone, instead of both particles $B_1$ and $B_2$,
based on the following experimental results:  (1) although the sum
of the two jumps in the Hall signal corresponding to the two
magnetization reversals,  $\Delta V_{H1}+ \Delta V_{H2}$, remains
constant, the ratio, $\Delta V_{H1}/\Delta V_{H2}$, varies from
1.1 to 1.9 in different measurements; (2) The size of particle
$B_2$ is expected to be smaller than particle $B_1$, and the
location of particle $B_2$ is more off-center than particle $B_1$,
so the contribution of particle $B_2$ to the Hall signal, if
observable, would be smaller than that of particle $B_1$. From the
Hall signal, we estimate that the diameter of particle $B_1$ to be
about 10-15\,nm \cite{SizeEst}.

\begin{figure}
\includegraphics[width=7.5cm]{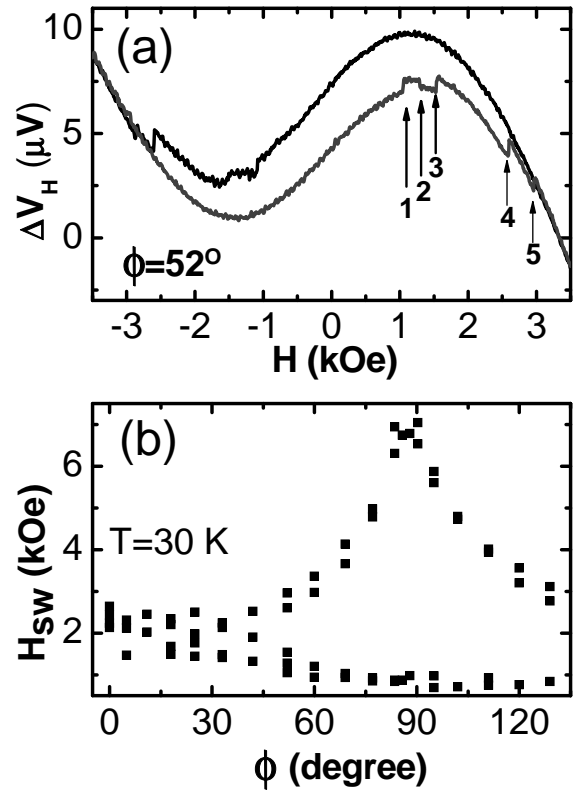}
\caption{\label{fig5} A typical hysteresis curve of particle $B_1$
obtained at $\phi=52$\,$^\circ$ and $T=30$\,K, which shows five
switching events as the field is swept up. (b) Angular dependence
of the switching fields of the same particle at $T=30$\,K.}
\end{figure}

The inhomogeneous nature of the particle is further evidenced by
the measurements in tilted fields.  Fig.\,4(b) shows part of four
hysteresis loops taken at exactly the same conditions
($\phi=18$\,$^\circ$ and $T=30$\,K). Among them, two loops have
four switching events ($\alpha_1$, $\alpha_2$, $\beta$, and
$\gamma$), while the other two only have three ($\alpha$, $\beta$,
and $\gamma$). Similar stochastic switching behavior has also been
observed at other field tilting angles. The number of switching
events for each magnetization reversal varies from two to as many
as five [Fig.\,5(a)]. The angular dependence of the switching
fields for a typical field sweep is plotted in Fig.\,5(b). The
distribution of switching fields shows a very complex behavior.
Nonetheless, the switching fields at high angles
($\phi>50$\,$^\circ$) are divided into two distinctive groups:  a
soft group (with lower switching fields) having a weak angular
dependence and a hard group with a strong angular dependence
similar to that for a typical single-domain elongated Fe
nanoparticle such as particle $A_1$. The multiple switching events
and their angular dependence suggest that the soft portion of
particle $B_1$ is probably composed of interacting clusters with
varying easy axes. The magnetic structure of this particle can be
modified by changes in the field sweep history, which is
illustrated in Fig.\,6: As the field is swept between $-5$\,kOe
and 5\,kOe, switching of both the soft and the hard portions takes
place (middle curve). On the other hand, the upper and the lower
hysteresis curves correspond to field sweeps between $-2$\,kOe and
2\,kOe, in which only the magnetization of the soft portion
switches. These two curves were taken under the same conditions
except for the starting field: the upper and lower curves were
obtained after the field was ramped to $-5$\,kOe and 5\,kOe,
respectively. The overall features of the two hysteresis curves
are very similar, however, neither is symmetric with respect to
$H=0$. They are shifted in opposite directions in $H$, resulting
in a large difference ($\sim0.3$\,kOe) in the switching fields.
The observed large difference in $H_{sw}$ of the soft portion can
clearly be attributed to the difference in the magnetization
direction of the particle's hard portion. Accordingly, the
configuration of the domain wall(s) between the soft portion and
hard portion is varied due to different field sweep history, which
in turn has a large effect on the switching of neighboring domains
through the exchange interaction. It should be noted that multiple
switching jumps have also been observed in electrodeposited Ni
\cite{Wernsdorfer96} and permalloy \cite{Sokolov02} nanowires with
$d=30$-90\,nm to which the curling mode would be applicable if the
nanowires were perfect.

\begin{figure}
\includegraphics[width=7.5cm]{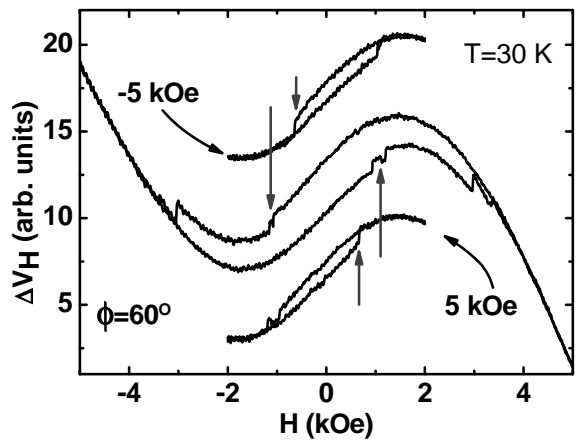}
\caption{\label{fig6} Three hysteresis loops of particle $B_1$
with different field sweep histories. The middle curve has a
history $-5$\,kOe$\rightarrow$ 5\,kOe$\rightarrow$ $-5$\,kOe, in
which both the soft and the hard portions of the particle are
switched. The upper and lower curves correspond to
($-5$\,kOe)$\rightarrow$ $-2$\,kOe$\rightarrow$
2\,kOe$\rightarrow$ $-2$\,kOe, and (5\,kOe)$\rightarrow$
2\,kOe$\rightarrow$ $-2$\,kOe$\rightarrow$ 2\,kOe, respectively,
in which only the magnetization reversal of the soft portion
occurs. }
\end{figure}

The existence of domain walls in particle $B_1$ and its absence in
particle $A_1$ as well as Fe nanowires of larger size prepared by
electrodeposition \cite{Sorop03} can be attributed to the growth
process. The growth of particle $B_1$ was performed with lower
precursor pressure and its growth rate was only about 1/4 of that
of particle $A_1$.  The growth of the Fe nanoparticles is
sustained by maintaining a constant tunneling current (typically
50\,pA), so a slower growth rate is likely to produce more
structural defects since the growth process suffers more external
(mechanical and electronic) interference. Some of these
imperfections provide the pinning forces for the domain walls.
Depending on temperature, applied fields, and the strength of each
pinning center, a domain wall may or may not be pinned down at a
particular place, resulting in stochastic behavior of the
magnetization reversal. Furthermore, the larger diameter of
particle $B_1$ may also be responsible for the observed
multi-domain structure due to the less dominant exchange
interaction. Finally, we speculate that there is a slight
possibility that there exists some degree of surface oxidation
which would provide extra pinning forces and nucleation sites for
particle $B_1$\cite{Kodama96}. These surface interactions might
also provide a possible explanation for the observed asymmetry in
the hysteresis loops at low temperatures. A quantitative analysis
of these effects, however, requires detailed information on the
microstructure of the nanoparticles, which is not possible with
our current experimental setup.

\section{Conclusion}

In summary, two types of elongated Fe nanoparticles with diameter
close to the coherence diameter have been studied in detail. One
has a single domain static state, but the magnetization reversal
cannot be described with coherent rotation. The data suggest a
thermally activated nucleation and propagation process, in which
possible nucleation sites include the ends and imperfections of
the nanocylinder. The other type of Fe particles studied in this
work has a multi-domain structure and shows complicated
magnetization reversal behavior. We have demonstrated that the
domain structure in these Fe nanoparticles can be manipulated
through the magnetic field history. The capability of submicron
Hall magnetometry in detecting the magnetization reversal of
single domains on the sub-10\,nm scale demonstrated in this work
is promising for important applications in understanding the
fundamental physics at dimensions comparable to magnetic exchange
lengths, as well as nascent fields such as bio-sensing and
non-invasive detection of spin-polarized carriers injected into
semiconductors.

\section{Acknowledgements}

Y.\ L.\ gratefully thanks R.\ Kallaher, S.\ Wirth and A. Anane for
technical help. This work was supported by NSF grant \#DMR0072395,
by DARPA through ONR grants \#N-00014-99-1-1094 and
\#MDA-972-02-1-0002, and by FSU Research Foundation through a PEG
grant. The work at Tohoku University was supported partially by a
Grant-in-Aid from the Ministry of Education, Japan, and by the
Japan Society for the Promotion of Science.


\begin{thebibliography}{}
\bibitem{Exlength}
  The definition of the magnetic exchange length is varied by a numerical factor in some literature.
\bibitem{SizeEst}
   A more accurate estimate of the diameter of particle $B_1$ is very difficult because this
   particle is located very close to the corner of the Hall cross.
\end{thebibliography}

\end{document}